\documentstyle[twocolumn,prb,aps,floats,epsfig]{revtex}

\begin{document}
\draft
\twocolumn[\hsize\textwidth\columnwidth\hsize\csname@twocolumnfalse\endcsname

\title{Pressure dependence of phase transitions in the quasi one-dimensional
       metal-insulator transition system $\beta$-Na$_{1/3}$V$_2$O$_5$}
\author{G. Obermeier, D. Ciesla, S. Klimm, and S. Horn}
\address{Universit\"{a}t Augsburg, 86135 Augsburg, Germany}
\date{\today}
\maketitle

\begin{abstract}
The pressure dependence of phase transitions in the quasi one-dimensional
vanadium oxide $\beta$-Na$_{1/3}$V$_2$O$_5$ has been studied by magnetic
susceptibility and electrical resistivity measurements. The pressure dependence
of the various transition temperatures is quite differently. The transition at
$T=240$~K, previously reported and attributed to ordering on Na sites, and a
second transition at $T \approx 222$~K, reported here for the first time and
attributed to a further increase of order on Na sites, are almost independent
of pressure. On the other hand, the metal-insulator (MI) transition at
$T_{MI}=130$~K shifts to lower temperatures, while the magnetic transition at
$T_N=24$~K shifts to higher temperatures with increasing pressure. We discuss
the different pressure dependencies of $T_{MI}$ and $T_N$ in terms of
increasing interchain coupling and the MI transition to be of Peierls type.
\end{abstract}

\pacs{PACS numbers: 71.30.+h, 62.50.+p}

\vskip3.0pc]

\narrowtext

\section{Introduction}
Quasi one-dimensional (1D) metals exhibit unique electronic properties, which
show potential for applications and, at the same time, for an expansion of the
understanding of electronic correlation effects in solids. The inorganic 1D
metal $\beta$-Na$_{1/3}$V$_2$O$_5$, which exhibits a metal-insulator (MI)
transition as a function of temperature at $T_{MI}=130$~K \cite{Yamada99} is a
system ideally suited to study the interplay between structural instabilities
and electronic correlations. This compound crystallizes in a monoclinic
structure exhibiting three inequivalent vanadium sites
\cite{Wadsley55,Khama89}. Two of the sites (V1, V2) have an octahedral and one
(V3) a five fold square pyramidal oxygen coordination. The latter site
resembles the V$^{5+}$ site in V$_2$O$_5$, which suggests a V$^{5+}$ occupation
also in Na$_{1/3}$V$_2$O$_5$. The oxygen octahedra of the V1 sites form a
zigzag chain in the b direction with octahedra sharing edges, while the oxygen
octahedra of the V2 sites form a double chain in b direction with octahedra
sharing corners along the b direction as well as in the ac-plane. It should be
pointed out that the V1 octahedra also share edges with the V2 octahedra and
that, in fact, the shortest V-V-distance is that between a V1 and a V2 site
\cite{Wadsley55,Khama89}.

This V-O framework provides tunnels along the b direction in which Na sites
form a double chain parallel b. For $x=1/3$ in Na$_x$V$_2$O$_5$ only half of
the sites are in fact occupied by Na ions, which build zigzag chains within
each tunnel. At room temperature there exists no long range correlation between
the zigzag chains. Below $T=230$~K X-ray diffraction (XRD) measurements
\cite{Yamada99} show, that the Na chains develop a long range order.

The intercalated Na atoms donate their valence electron to the host lattice,
which results in a mixed valence V-state with one $3d$-electron per six
vanadium ions. The distribution of the $3d$-electrons on the three different
vanadium sites is not completely clear. Different electron configurations in a
charge ordered state below $T_{MI}$ have been discussed
\cite{Itoh00,Itoh01,Nishimoto01,Ueda01}. In contrast to related systems, like
e.~g.\ $\alpha^\prime$-NaV$_2$O$_5$ and $\eta$-Na$_{1.29}$V$_2$O$_5$, which are
structurally low dimensional and electrically insulating,
$\beta$-Na$_{1/3}$V$_2$O$_5$ is a quasi one-dimensional metal at room
temperature. In resistivity measurements above $T_{MI}$ metallic behavior is
observed parallel to the b axis, while perpendicular to b semiconducting
behavior occurs with an anisotropy ratio of $\rho_\perp/\rho_\parallel > 100$.
At $T_N=24$~K a magnetic transition occurs \cite{Yamada99}, which has been
shown by NMR measurements to be long range in nature \cite{Itoh00}. From EPR
and magnetization measurements \cite{Ueda01,Vasil01,Schlenker79} it was
concluded, that the ordered state is a canted antiferromagnet, with the
ferromagnetic component aligned along the b axis. Large differences between
zero field and field cooled magnetization measurements are observed
\cite{Yamada99}. Such ''spin glass like'' behavior has been observed before in
systems with magnetically strongly anisotropic behavior and was attributed to
magnetic domain effects \cite{Tsurkan01}. The quasi 1D structure of
$\beta$-Na$_{1/3}$V$_2$O$_5$ suggests that such a magnetic anisotropy exists in
this system.

The electronic transport and magnetic properties are very sensitive to the
Na-stoichiometry $x$. Metallic behavior is observed only in a very narrow range
of Na-concentration and a slight deviation from $x=1/3$ induces semiconducting
behavior in the electrical resistivity, most likely due to disorder effects.
Deviation from the exact stoichiometry also rapidly suppresses the magnetic
transition \cite{Yamada99}. External pressure is expected to change the
relative strength of intra- to interchain interactions and therefore can yield
information about the influence of the dimensionality of the system on the
different phase transitions. In this paper we report on the effect of external
pressure on the phase transitions in $\beta$-Na$_{1/3}$V$_2$O$_5$ studied by
magnetic susceptibility and electrical resistivity measurements.

\section{Experimental}

Needle like single crystals of a typical size of $4\times0.4\times0.2$~mm$^3$
were grown by a flux method similar to that described in
Ref.~\onlinecite{Yamada99}. Laue diffraction showed the longest dimension was
parallel to the crystallographic b axis. For resistivity measurements along b
four gold pads were evaporated on the crystals. Then 40~$\mu$m platinum wires
were attached to those pads by silver epoxy. The resistivity was measured using
a standard four-probe DC-technique, i.~e.\ for each measurement the current was
reversed to eliminate voltage offsets. AC-susceptibility measurements were
performed using a mutual inductance bridge (MIB) in which several crystals were
aligned with the b axis parallel to the magnetic field. The amplitude of the
field was about 1~G at a frequency of about 1~kHz. Quasi hydrostatic pressure
experiments were carried out in a self clamped CuBe piston-cylinder cell with
n-pentane -- isoamyl (1:1) as a pressure transmitting medium. The pressure was
measured utilizing a manganin gauge in case of the resistivity measurements and
the pressure dependence of the superconducting transition of lead in case of
the AC-susceptibility. DC-magnetic measurements at ambient pressure were
performed using a Quantum Design SQUID magnetometer.

\section{Results}

\subsection{Electrical resistivity}

\begin{figure}[bt]
 \epsfig{file=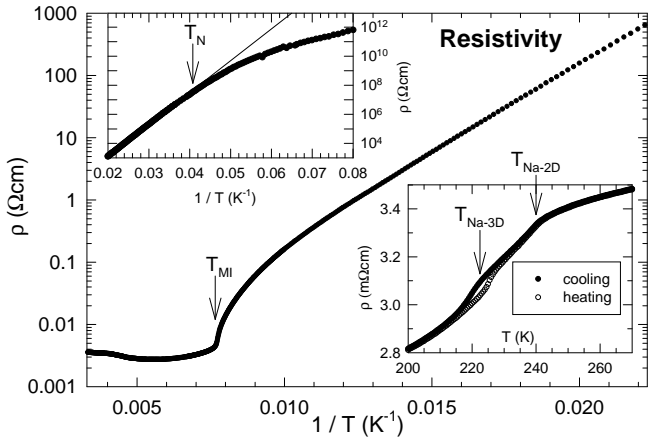, width=\columnwidth, clip}
 \caption{Arrhenius plot of resistivity. The right inset shows the high
  temperature resistivity on a linear scale. The left inset shows the low
  temperature resistivity measured by a two-probe method.}
 \label{rho}
\end{figure}

The temperature dependence of the b axis resistivity of a nominally
stoichiometric Na$_{1/3}$V$_2$O$_5$ sample is shown in Fig.~\ref{rho}.
A steep increase in resistivity marks the MI transition at $T_{MI}=130$~K.
Below 80~K the resistivity is characteristic of a semiconductor with an
activation energy of 0.05~eV. The left inset of the figure shows the extension
of the resistivity from 50~K down to lowest temperatures measured in a special
cryostat with uninterrupted triaxial wiring terminating just at the sample. A
deviation from linearity in the Arrhenius plot towards lower resistivity is
observed below the magnetic transition at 24~K.

The right inset of the figure shows the metallic behavior at high temperatures.
At 240~K, denoted as $T_{Na \mbox{-} 2D}$, the slope of resistivity changes
rapidly, which was attributed to ordering between Na-chains \cite{Yamada99}.
Below 230~K another noticeable anomaly in the resistivity marks a second
transition, which shows a hysteresis upon thermal cycling and which we also
attribute to a structural transition, associated with further ordering of the
Na sublattice, as discussed below. The center of the hysteretic transition is
located at $T_{Na \mbox{-} 3D}=222$~K.

Due to the strongly anisotropic resistivity of Na$_{1/3}$V$_2$O$_5$ an accurate
alignment of current and voltage contacts is essential to avoid contributions
from non b axis components of the resistivity. Therefore, the negative slope of
resistivity just above $T_{MI}$ might rather reflect a non ideal position of
contacts than a deviation of the b axis resistivity from metallic behavior in
this temperature range. Some measurements on samples of the same batch as the
sample shown in Fig.~\ref{rho} display a maximum of resistivity at 240~K and
semiconducting behavior above, which we also attribute to contributions other
than the b axis component of the resistivity tensor, which show semiconducting
behavior as a function of temperature.

Slightly off-stoichiometric samples displayed a negative non-exponential
temperature characteristic over the whole temperature range, most likely due to
disorder. Owing to the quasi one-dimensional nature of the compound such a
sensitivity of the transport properties to disorder can be expected. The jump
in resistivity at 130~K, marking the MI transition, however, is still observed
in such samples.

The transitions observed at $T_{Na \mbox{-} 2D}=240$~K, $T_{MI}=130$~K and
$T_N=24$~K have been reported previously, while the transition at $T_{Na
\mbox{-} 3D}=222$~K is reported here for the first time. Since we observed the
latter transition, unresolved in earlier studies, in all of our stoichiometric
samples we consider it to be an intrinsic property of Na$_{1/3}$V$_2$O$_5$.

\subsection{Magnetic susceptibility}

\begin{figure}[bt]
 \epsfig{file=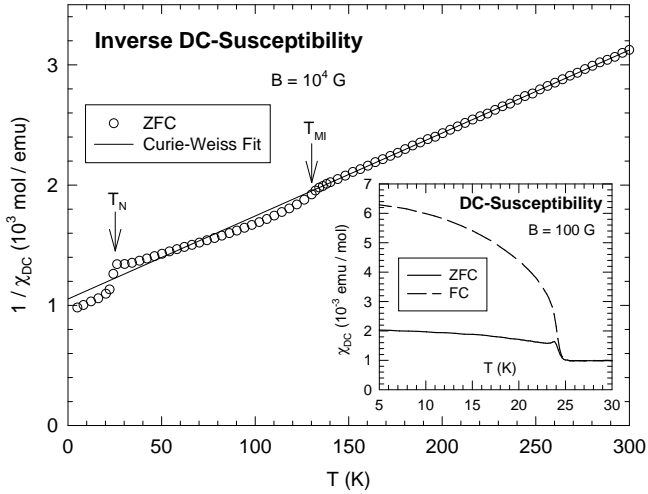, width=\columnwidth, clip}
 \caption{Inverse magnetic DC-susceptibility of single crystals as a function
 of temperature measured in a magnetic field of $B=10^4$~G parallel to the b
 axis. The inset shows the magnetic susceptibility when zero field cooled
 (ZFC) and field cooled (FC) measured with $B=100$~G.}
 \label{chi_DC}
\end{figure}

Fig.~\ref{chi_DC} shows the inverse magnetic DC-susceptibility as a function of
temperature measured in a field of $10^4$~G along the b axis.
The susceptibility follows a Curie-Weiss law in the temperature range
150~K~$<T<$~300~K with a Curie-Weiss temperature $\Theta_{CW}=-150$~K and an
effective moment of $1.9~\mu_B$ per V$^{4+}$ ion. This is in agreement with
literature \cite{Yamada99,Schlenker79,Goodenough70} where values of $\mu_{eff}$
in the range $1.9 < \mu_{eff} < 2.2~\mu_B/V^{4+}$ have been reported.

The fact that the magnetic susceptibility does not show a noticeable anomaly at
the phase transitions at $T_{Na \mbox{-} 2D}=240$~K and $T_{Na
\mbox{-}3D}=222$~K, indicates that the magnetic and electronic state of the
vanadium ions is not affected by these transitions. In contrast, at the MI
transition a sharp upturn of the susceptibility is observed. The onset of
magnetic order at 24~K is reflected by a sharp rise of $\chi$ (as measured in a
field of $10^4$~G). At lower magnetic field this rise becomes more pronounced
and for zero field cooled curves a small cusp is observed at $T_N=24$~K, as
shown in the inset of Fig.~\ref{chi_DC}.

\subsection{Pressure dependence}

\begin{figure}[bt]
 \epsfig{file=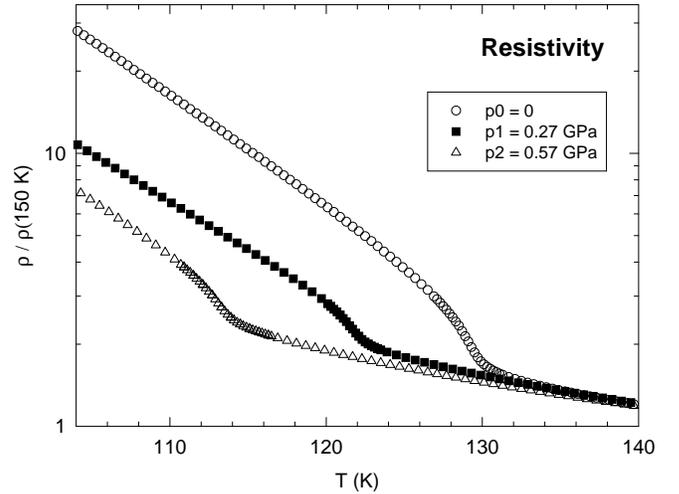, width=\columnwidth, clip}
 \caption{Resistivity normalized to the value at $T=150$~K as a function of
  temperature in the vicinity of $T_{MI}$ at different pressures.}
 \label{rho_p}
\end{figure}

The difficulties to measure the clean b axis contribution of resistivity are
even more complex in a high pressure environment. Therefore, a measurement of
the absolute resistivity was not possible under pressure. Nevertheless, the
characteristic features marking the various phase transitions are easily
resolved. The two transitions at $T_{Na \mbox{-} 2D}$ and $T_{Na \mbox{-} 3D}$,
are independent of pressure within the experimental resolution. This suggests,
that there is a link between these two transitions. The resistivity as a
function of temperature in the vicinity of $T_{MI}$ is shown in
Fig.~\ref{rho_p} for three different pressures, as indicated in the figure. In
contrast to the transitions at higher temperatures, the MI transition reacts
extremely sensitive to pressure and shifts to lower temperatures at a rate of
28~K/GPa.
Qualitative information of the pressure dependence of the ratio of the
resistance parallel b to the resistance perpendicular b was obtained from
measurements on one sample with six contacts attached. We observe a decrease of
anisotropy of resistivity with increasing pressure, indicating a crossover from
1D to two- or three-dimensional behavior, presumably due to increased
interaction between vanadium chains.

\begin{figure}[bt]
 \epsfig{file=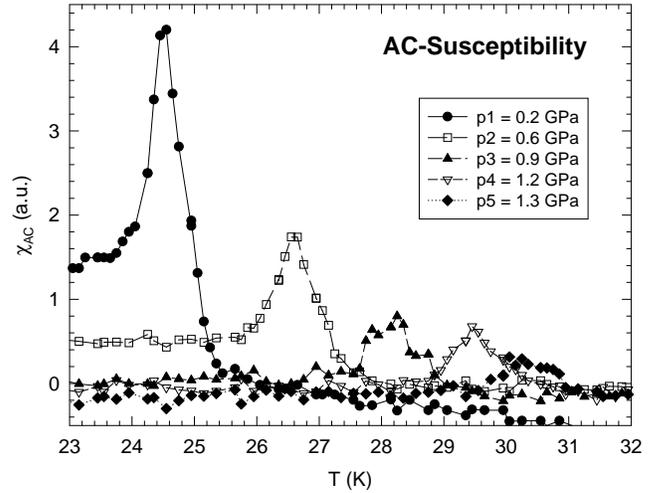, width=\columnwidth, clip}
 \caption{Inductive signal related to the AC-susceptibility as a function of
  temperature in the vicinity of $T_N$ at different pressures.}
 \label{chi_p}
\end{figure}

To study the pressure dependence of the magnetic transition temperature $T_N$,
we performed AC-susceptibility measurements using a miniature MIB. The signal
obtained from the MIB does not allow to evaluate the absolute value of
$\chi_{AC}$. The magnetic transition, however, is clearly seen as a cusp in the
signal (Fig.~\ref{chi_p}) similar to the run of the ZFC DC-susceptibility curve
(Fig.~\ref{chi_DC}). With increasing pressure $T_N$ shifts to higher
temperatures at a rate of 5~K/GPa while the height of the cusp decreases.

\begin{figure}[bt]
 \begin{center}
  \epsfig{file=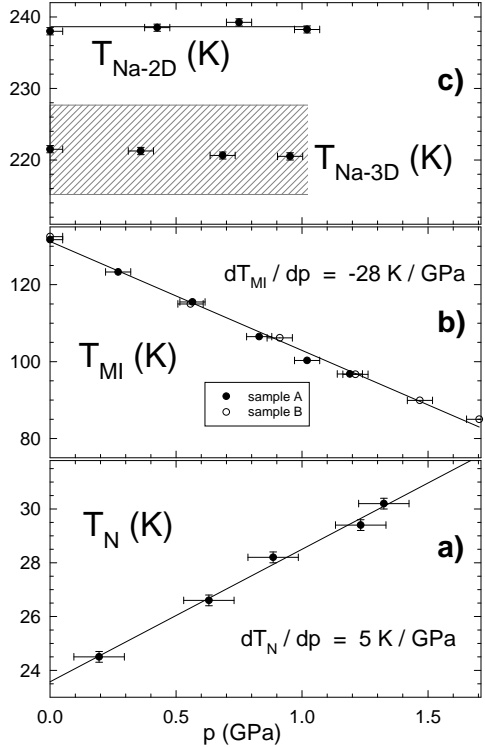, width=6.5cm, clip}
 \end{center}
 \caption{Pressure dependencies of the various phase transitions as determined
  by resistivity or susceptibility measurements. Please note the different
  temperature scales for (a) $T_N$, (b) $T_{MI}$, and (c) the transitions
  related to the ordering of the Na sublattice. The shaded area in panel (c)
  marks the region of hysteresis.}
 \label{T_p}
\end{figure}

Fig.~\ref{T_p} summarizes the pressure dependencies of the respective phase
transitions.
$T_{MI}(p)$ was investigated for two different samples (open and filled symbols
in Fig.~\ref{T_p}(b)) both showing the same pressure dependence as expected. A
linear extrapolation to higher pressures would predict the magnetic and the MI
transition to coincide at $p=3.2$~GPa.

\section{Discussion}
As already mentioned above, the pressure independence of the resistivity
anomaly at $T_{Na \mbox{-} 2D}=240$~K and the hysteretic phase transition at
$T_{Na \mbox{-} 3D}=222$~K implies a common origin. XRD measurements by Yamada
et al. revealed satellite reflections below 230~K, indicating a doubling of the
unit cell in b direction \cite{Yamada99}. From a comparison to the intensity of
similar satellite reflections found for $\beta$-Ag$_{1/3}$V$_2$O$_5$ Yamada et
al. deduced that the reflections are due to ordering on the Na (Ag) sublattice.
The steep decrease of resistivity below $T=240$~K was then attributed to this
ordering. However, the reported temperature below which the satellite peaks
appear ($T<230$~K), coincides rather with the hysteretic transition in
resistivity observed here (see right inset of Fig.~\ref{rho}). We therefore
attribute this hysteretic transition to ordering within the Na sublattice and
suggest the following scenario: at room temperature the Na ions form regular
zigzag chains along the b axis. Within each zigzag chain the Na ions have a
periodicity of 2b. Perpendicular to the b direction these zigzag chains are not
correlated with respect to their occupation, and, therefore, the probability of
occupation of each Na site is one half on the average and the lattice parameter
is b. Below $T_{Na \mbox{-} 2D}=240$~K correlations regarding the occupation of
Na sites between neighboring Na zigzag chains develop, resulting in the steep
decrease of resistivity. Since the distance between Na chains is shorter along
the a ($d_a \approx 10$~\AA) than along the c axis ($d_c \approx 15$~\AA),
correlations probably will first develop in the a direction. This results in
long range correlations of Na chains within (ab-)planes, while there are no (or
only short range) correlations perpendicular to these planes. Still the average
occupation number of all Na sites remains one half and the periodicity of the
lattice remains unchanged. Finally, below $T_{Na \mbox{-} 3D}$ a coherent phase
relation between the 2D-ordered chains emerges, which results in a transition
to a three-dimensional long range ordered state associated with a hysteresis in
the resistivity measurements and the appearance of satellite reflections
signaling a doubling of the lattice parameter along the b direction.

The phase transitions at $T_{MI}=130$~K and $T_N=24$~K show a pronounced
pressure dependence and clear signatures in the magnetic susceptibility,
indicating that the electronic and magnetic state of the V ions changes at the
respective transition. According to NMR measurements \cite{Itoh00} the
$3d$-electrons occupy the V1 and V2 sites above $T_{MI}$. Since both sites form
chains in b direction (see above), the resulting electronic structure is
expected to exhibit quasi one-dimensional features. This expectation is
confirmed by band structure calculations based on density functional theory,
which show a pronounced dispersion parallel, but a very small dispersion
perpendicular to the b axis \cite{Eyert}. This one-dimensional nature of the
electronic structure and its concomitant Fermi surface nesting will leave
Na$_{1/3}$V$_2$O$_5$ susceptible to a Peierls like transition. The pressure
dependence of the MI-transition is indeed consistent with such a scenario. In
principle, a one-dimensional system is dominated by fluctuations and no phase
transition is expected for $T > 0$~K. In reality there is always a finite
coupling between the one-dimensional objects constituting the system. This
finite coupling results in a phase transition at finite temperature. Assuming
that the MI transition is associated with Fermi surface nesting, the shift of
the transition temperature $T_{MI}$ to lower temperature with pressure can be
understood as follows: application of pressure will increase the coupling
between vanadium chains and, at the same time, the lattice stiffness. The
increase of interchain interaction will increase the $3d$-band dispersion
perpendicular to the chains and, concomitantly, reduce the nested portion of
the Fermi surface and, consequently, $T_{MI}$ decreases. An increase of lattice
stiffness increases the amount of strain energy necessary for the lattice
distortion associated with a Peierls like transition, decreasing $T_{MI}$
further. The scenario of an increased interchain interaction with pressure is
supported by the evolution of the magnetic transition temperature with
pressure. Increased interchain interaction will suppress fluctuations
characteristic of the quasi one-dimensional system and shift the magnetic
transition to higher temperatures. At the same time, this leads to a reduction
in magnitude of the magnetic signature with increasing pressure, since the
three-dimensional freezing of the spin structure occurs already at higher
temperatures, i.~e.\ smaller intrachain correlation lengths.

The pressure dependence of both $T_{MI}$ and $T_N$ therefore are consistent
with the Peierls like scenario described above. To put forward an explanation
for the fact that such a transition results in an insulating state one has to
look at the electronic structure of Na$_{1/3}$V$_2$O$_5$. The V $3d$ states
associated with both the V1 and V2 sites (but not with the V3 site) appear to
be occupied above $T_{MI}$ \cite{Itoh00}. Accordingly, one $3d$-electron is
shared between two sites, i.~e.\ there is one electron for four vanadium atoms.
It also follows that the V $3d$ groundstate energy levels of V1 and V2 sites
are (almost) degenerate, resulting in a 1/8 filling of each state above
$T_{MI}$. To achieve an insulating ground state (and remove the degeneracy) the
lowest lying levels have to be split. Below $T_{Na \mbox{-} 3D}$ the doubling
of the unit cell in b direction and the concomitant band splitting will lead to
1/4 filled bands. Below $T_{MI}$ further splitting can occur due to the
formation of a (commensurate) charge density wave with suitable propagation
vector q below $T_{MI}$. Depending on the propagation vector q this may result
in a completely filled band separated from the other bands by a band gap, or in
a half filled band. In the latter case an insulating state may follow from
strong on-site Coulomb interaction. The NMR results \cite{Itoh00} suggest a
charge ordering with the $3d$-electrons either completely on V1 or V2 sites. In
a local picture, consistent with the Peierls like scenario described above,
this can be realized by a dimerization of either V1 or V2 sites, resulting in a
bonding-nonbonding splitting of the dimerized levels and the occupation of the
bonding state. The resulting half filled band situation again could result in
an insulating state due to on-site Coulomb interactions. Clearly these
suggestions are highly speculative and more investigations on the electronic
structure and the lattice distortions at the MI-transition have to be carried
out to obtain a better understanding of the insulating state and the
distribution of the $d$-electrons on the various V sites in the metallic and
insulating state of Na$_{1/3}$V$_2$O$_5$.

\section{Conclusion}
The recorded pressure dependencies of the transition temperatures of the MI
transition and the magnetic transition in Na$_{1/3}$V$_2$O$_5$ are consistent
with the picture of a quasi one-dimensional system. The decrease of the MI
transition temperature with increasing pressure is interpreted as a result of a
loss of Fermi surface nesting due to increased interchain interactions and the
concomitant suppression of a Peierls like transition. At the same time the
increased interchain interaction causes an increase of the magnetic ordering
temperature. An observed pressure independent hysteretic transition is
attributed to long range three-dimensional ordering on the half filled Na
sublattice.

\acknowledgments

This work was supported by the Deutsche Forschungsgemeinschaft under Contract
Nos. HO~955/2-1 and SFB 484. We are grateful to P. Riseborough for valuable
discussions.

\end{document}